# Emotional Qualities of VR Space

Asma Naz[1]
University of Texas at Dallas, USA

Regis Kopper[2]
Duke University, USA

Ryan P. McMahan[3]
University of Texas at Dallas, USA

Mihai Nadin[4]
University of Texas at Dallas, USA

**ABSTRACT**

The emotional response a person has to a living space is predominantly affected by light, color and texture as space-making elements. In order to verify whether this phenomenon could be replicated in a simulated environment, we conducted a user study in a six-sided projected immersive display that utilized equivalent design attributes of brightness, color and texture in order to assess to which extent the emotional response in a simulated environment is affected by the same parameters affecting real environments. Since emotional response depends upon the context, we evaluated the emotional responses of two groups of users: inactive (passive) and active (performing a typical daily activity). The results from the perceptual study generated data from which design principles for a virtual living space are articulated. Such a space, as an alternative to expensive built dwellings, could potentially support new, minimalist lifestyles of occupants, defined as the *neo-nomads*, aligned with their work experience in the digital domain through the generation of emotional experiences of spaces. Data from the experiments confirmed the hypothesis that perceivable emotional aspects of real-world spaces could be successfully generated through simulation of design attributes in the virtual space. The subjective response to the virtual space was consistent with corresponding responses from real-world color and brightness emotional perception. Our data could serve the virtual reality (VR) community in its attempt to conceive of further applications of virtual spaces for well-defined activities.

**Index Terms**: I.3.7 [Computer Graphics]: Three-Dimensional Graphics and Realism—Virtual Reality

## 1. INTRODUCTION

This paper reports on an experimental study of virtual environments (VEs) explored as an architectural tool to evaluate a new design concept of a real-world dwelling for information-age professionals, also known as *neo-nomads*. The severe housing crisis of the Silicon Valley in recent years has led architects and interior designers to search for new concepts of compact living to accommodate the transitional lifestyles of these technology-dependent, mobile professionals [19]. The dwelling envisioned in this paper consists of a single living space with variable design parameters of color, light and texture used to generate multiple emotional experiences of spaces that could potentially support daily living of occupants. In a user study, a six-sided immersive Cave Automatic Virtual Environment (CAVE)-type display was used to simulate the living space that adapted equivalent design parameters of color, brightness and texture to create a set of emotional experiences of spaces. The study measured perceived emotional dimensions of virtual spaces in nine categories: warmth, coolness, excitement, calmness, intimacy, spaciousness and comfort, as well as spatial preferences for two activities: work and rest. Since emotional responses to space depend on context, the influence of activity on perception was also assessed by evaluating responses from two groups of users: an active group that performed a typical daily activity and an inactive (passive) group.

The study presented a hypothesis that the real-world phenomenon of affective space creation of traditional architecture could be replicated in a simulated environment through modification of design parameters of color, brightness and texture. The study goal was to assess the extent to which design parameters influenced emotional responses to virtual spaces. Subjective responses in the VE were analyzed to find correlation with corresponding normative real-world emotional space perception related to color, brightness and texture.

The outcome of the study is in the form of data relevant for design of a real-world dwelling that has the capacity to generate variable spaces with affective dimensions in order to support minimalist lifestyles for young professionals. The major contribution of this study lies in its definition of virtual reality (VR) as a viable evaluation platform for new architectural concepts where variable design parameters can be adapted to create experiential spaces with emotional or psychophysiological qualities. Such data could have real-world implications in decision-making processes for architects, builders, interior designers and clients, as well as facilitate better communication in terms of time and resources.

## 2. RELATED WORK

In this section, we review prior literature on VR in the context of architecture and affective dimensions, and also provide an overview of affective design parameters for living spaces.

### 2.1 VR as Architectural Design Evaluation Tool

Virtual Reality technology has been used as an evaluation tool in architecture, mainly to assist in decision-making during design process to explore unbuilt concepts and difficult-to-realize projects, as well as communal, educational or recreational designs that need participatory inputs from architects, designers, clients and users in a collaborative setting [7]. Research on architectural applications of VR was initially envisioned by Sutherland [24] as virtual reality consisting of realistic living spaces. The first head-mounted display (HMD), designed by Sutherland, had an architectural space as one of the first immersive VEs [24] followed later by the first room-sized VR—CAVE [23], with four 10-foot projection screens and interactive user interfaces. In CAVE, a prototyping system called CALVIN introduced multiple perspectives for architectural design at both ground and global (above-ground) levels [16].

In VR architectural studies, walk-throughs allow users to travel freely in the VE, and prototyping systems provide user-centric designs during early development phases [7]. Two design studies during the mid-1990s conducted walk-through studies to evaluate architectural spaces for unnecessary design elements [17]. In another study conducted at the University of Washington, working and testing design methods in VR improved design quality of students [17]. In the late 1990s, Fröst and Warren used a more sophisticated, immersive CAVE system with multiple participants and real-time image generation to compare traditional paper and pencil design sessions with VR technology for the initial conceptual and schematic design process in architecture [12]. Results showed that VR proved to be a better tool to conceptualize, analyze, test and construct architectural designs. Over the years, the benefits of VR have been acknowledged in the architectural design process as an engaging, interactive system with a high degree of realism [17].

---
[1] asma.naz@utdallas.edu
[2] regis.kopper@duke.edu
[3] rymcmaha@utdallas.edu
[4] nadin@utdallas.edu

## 2.2 VR Studies in Measuring Affective Dimensions

Assessment of spatial quality of architecture in VR has involved the measurement of both qualitative and quantitative factors of architectural space design. Quantitative measures are mostly related to size, proportions, scale and distance perceptions, whereas qualitative factors are affective and environmental aspects of space. To quantify emotional and environmental preferences, some basic models were established in environmental psychology [9]. These models used semantic differential scales (non-numerical scales) for affective appraisals of space constructed with pairs of oppositional adjectives as descriptors of environmental aspects of mood [14]. The adjectives are part of everyday language that convey both emotional and physical meanings to characterize perceived architectural space dimensions. These adjectives relate to size, openings, temperature, color brightness, style, form and structural details of architectural spaces, such as: pleasant and unpleasant, bright and dull, arousing and calm, narrow and spacious, dark and bright, open and enclosed, and warm and cold [10].

Some studies conducted in VR have compared qualitative virtual perception to corresponding real-world perception in order to find out how similar these spaces feel [15]. These studies established quantitative relationships between perceived emotional aspects and simulated volume of space: scale, lighting, architectural details and other artifacts [15].

Attempts have been made to incorporate VR as part of architectural design curricula for design scheming, development, testing and review phases. One such example was the College of Architecture and Planning (CAP) at Ball State University, which used HMD-based VR to assist students understand spaces in terms of everyday semantics with affective dimensions and manipulate space-making features to create spaces with meaning [7]. Students reproduced real-world functional spaces in the VE to compare how they are similar in negative or positive appreciations [7]. Certain studies show that, similar to generic pictorial features, spatial features also elicit emotional responses [9]. Previous studies have used semantic scales to quantify and establish correlations between everyday language that describes perceived experiential qualities and certain spatial features of interior rooms: size, shape, dimension, color and saturation [11]. Another VR-based perceptual study with the *Elumens Vision Station* used a series of still images of interiors to establish relationship between physical openness (window, room size) and affective qualities of space [9].

As far as we could find in the literature, very few studies and 3D design tools have focused on the *qualitative* factors of architectural spaces. Not enough research has been conducted on sensory and psychological dimensions of space perception. Architectural design is inherently spatial, and space perception is a sensory experience that carries meanings with emotional dimensions. Affective architecture articulates color, light, materiality, form and texture to create sensory perceptive, visually engaging spatial experiences to elicit emotional responses from the occupant [19].

The use of VR technology is yet to be explored in the psychological and metaphysical dimensions of natural light, materiality, color and texture that bring subjective meanings, essences and sensuousness into space-making as an essential part of the architectural design process.

This paper explores the direct emotional response to the space design parameters—a phenomenon that we believe has not been explored in any previous VE studies related to affective architecture. The human-scale immersive CAVE is appropriate for this study for its large-scale simulations, high resolution, photorealistic real-time projections on all six screens, ability to use mixed reality elements, and 360-degree fully immersive view for an enhanced sense of presence for the viewer.

## 2.3 Affective Space Design Parameters

The visual space perceptual process depends on the properties of architectural elements—color, light, material, texture, size and shape—as well as their interrelationships [5]. Standard architecture and interior design practices use certain design principles to create affective dimensions of space. The affective aspects of spaces are psychophysiological dimensions of perceived spatial experience that elicit or influence an occupant's moods and emotions, feelings, preferences and attitudes. Subjective expression and interpretation of spatial expression are driven by context: culture, imagination, memory, thoughts or previous emotional states. Mood and experienced spatial environment are interdependent [10].

### 2.3.1 Color

Warm and cool are sensorial emotions related to hue [13]. Warm colors are at the red end of the color spectrum, consisting of red, orange, yellow and their combinations. Cool colors are at the opposite end of the spectrum, consisting of blue, green, purple and their combinations. Artists, architects, interior designers, scientists, physicians and psychologists have worked with the science of color—i.e., its impact on the human physiology, psyche and behavior, its natural and cultural associations, and interactions between colors. Color acts as an emotional, physical and physiological stimulant. Its perception is contextual. Studies have acknowledged certain affirmations regarding affective aspects of color that are related to arousal and dominance [11]. Warm colors feel arousing or exciting, while cool colors feel calming [11]. The causes behind emotional and physiological responses to a color are manifold and complex, yet interrelated. Perception of red, as a symbol of romance, passion, power, criticism, anger, warning or violence may stem from its natural or biological associations with health, blood, stamina, vigor, heat or flame [2]. Similarly, blue may soothe and calm as the color of the sky. (On the other hand, the idea of "feeling blue" as an expression of sadness may have biological roots with emotional and cultural associations.) Cultural connotations of a color can have strong impact on human attitude and behavior. Experiments have demonstrated that student engagement in class could be improved by avoiding the use of red in grading papers, and male aggression in jail cells could be reduced by painting interior walls pink, a color that is culturally recognized as feminine [2].

Color has an impact on perception of spatial depth as well [26]. Warm colors are known as "advancing colors" that appear to be closer, and cool colors are known as "receding colors" that appear to be farther away [13]. Artists and architects from the Bauhaus and De Stijl movements explored perception of pure colors, their complex interactions with geometric shapes and forms, and spatial experiences related to color [8]. Architect Le Corbusier conceived spaces with planar surfaces of pure and vibrant colors and examined their interactions with exposed materials to elicit emotional responses [6].

### 2.3.2 Light

Light and color are inseparable. Perception of color in an interior space depends on direct and indirect illumination, type of light and interactions between color and light [13]. Light can still appear "warm" or "cool" based on the feeling it evokes when perceived with color and space as a whole [13]. Additionally, light can provide materials with a sense of depth. Cool and light colors feel more spacious than warm and dark colors [11].

### 2.3.3 Texture

Texture is the feel, appearance and consistency of a surface [25]. In this paper, texture refers to the surface property of rendered materials that construct perceived space. Visual perception of spatial texture is informed by past tactile experience of material weight, shape and resistance [20]. The distinct character of spatial texture of a perceived space derives from the texture of each and all surrounding surfaces perceived as a whole [21]. Roughness and smoothness of texture are characteristics of texture graininess, varying in density, size, orientation and depth of grain. Affective qualities of perceived texture are inherited from material quality and temporality: age, origin, inherent permanence or impermanence, origin and construction process. A rough texture of a brick or stone may appear richer in character, exciting and "honest" in its origin, while smooth texture, such as plaster, may appear comparatively dull or boring [21]. Depending on material quality, darker or lighter color tone, the perceived character of texture may be hard or soft, heavy or light, warm or cool [21]. Depending on visual weight and tactility of material and its texture, a space may feel spacious or intimate, warm or cool.

## 3 USER STUDY

We performed a user study to assess affective responses to perceived spatial experiences. Participants were invited to a test session inside a simulated living space rendered in a 6-sided CAVE-type system. Each participant was shown a set of virtual spaces and was asked to rate them quantitatively through a set of questions. Half of the participants performed an activity during experiment, while the other half remained inactive.

### 3.1 Method

The study used a mixed design with three *within-subject* factors—color, brightness and texture—and one *between-subjects* factor—idling/performing an activity. The *within-subject* design was used to establish quantitative correlations between variable design parameters (independent variables) and emotional responses to spaces. The design parameters each had two levels: color (orange and blue), brightness (dark and light) and texture (rough and smooth). The dependent variables were the quantitative user ratings of affective dimensions of spaces. The *between-subjects* factor was used to assess the impact of activity on perception.

Participants were randomly distributed over two experimental groups: one which required participants to perform an activity (active) and one where participants remained idle (inactive). Table 1 summarizes the participant pool and distribution over the experimental groups. In the *active* condition, sixteen users performed a minimal-effort daily activity of folding real clothes from a laundry basket and piling them up on a real table. In the *inactive* condition, the users remained still.

Table 1. Between-subjects design of the user study

| Experimental Condition | Activity | Participants | Male/Female Ratio |
|---|---|---|---|
| Active | Yes | 16 | 9/7 |
| Inactive | No | 16 | 8/8 |

### 3.2 The Virtual Environment

#### 3.2.1 Space Layout and Apparatus

A single, square living space (3.35 m x 3.35 m x 3.35 m) was simulated in the Duke immersive Virtual Environment (DiVE), a projection-based, 6-sided CAVE-type display at Duke University. Screen resolution was 1920x1920. The living space was extended beyond the actual size of display (2.9m x 2.9m x 2.9m) to stimulate depth perception due to binocular disparity. The living space was enclosed by four walls, a ceiling and a floor. It consisted of a mixed reality (MR) setup, where both real and virtual furniture were rendered to provide a sense of scale and enhance presence. A study table and a small bed were simulated and placed abutting two opposing walls (figure 1). A real chair and a round table were placed inside the virtual room flanked with the simulated study table on its left and the bed on its right. A real laundry basket full of unfolded clothes was placed on the ground in front of the chair (figure 2).

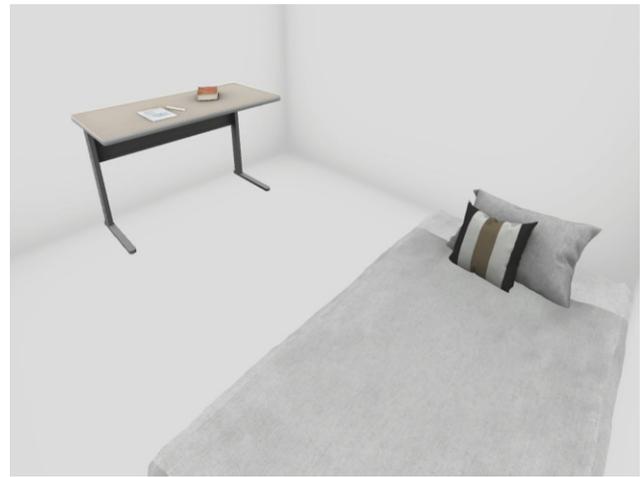

Figure 1: Simulated furniture in VE.

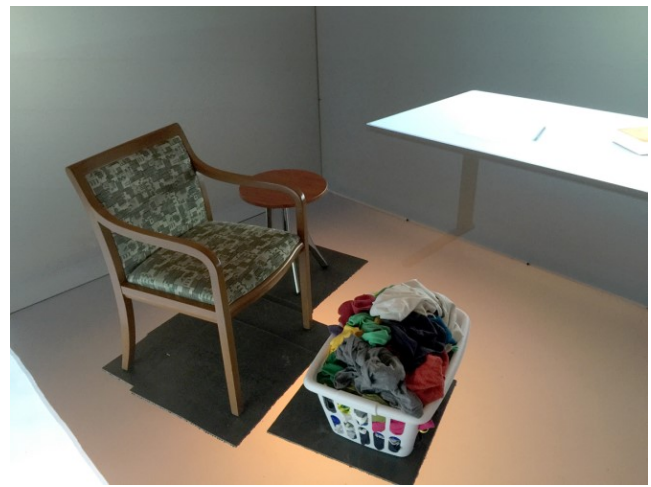

Figure 2: Real furniture and props placed inside virtual environment.

The chair was placed at a distance of approximately 210 cm from the facing wall. The door to the CAVE-type display was behind it. A calibrating gridline on the floor of the display was used to provide reference lines and mark the location of the real furniture and props for each test session. While the user was seated, eye height was approximately 1.25-m above the floor. No travel or other interaction was involved in the test, except for natural head movement while seated. Participants wore active stereo shutter glasses and sat on a chair throughout the testing session. An Intersense IS-900 ultrasonic tracker (head) was used to measure location and orientation of participants for proper computer simulation. This spatial setting, including size of space, furniture layout, configuration and the activity props remained constant across the two experimental conditions.

The concept of a multi-functional single space draws its inspiration from the efficiency or studio apartments, and "micro-apartments" of overcrowded, high-density big cities, such as San Francisco and New York [22]. Many tiny apartments rented by the digital age population have single, multi-functional living areas as small as 9.3 m$^2$ (100 sqft) [27]. These spaces accommodate overlapping activities with multipurpose, compact furniture and storage for minimalist belongings of neo-nomads [4].

### 3.2.2 Space Layout and Apparatus

Three types of texture maps were used for the simulated virtual surfaces of the living space in order to represent three commonly used materials. These were stone, drywall and carpet (fabric). A stone texture map was applied only on one wall (facing the user) that acted as the main accent wall of the room. The surrounding three walls and ceiling were rendered with drywall texture. A carpet texture was applied on the floor. All texture maps represented the actual scale of real materials (figure 3). The texture maps of the simulated furniture had a predominantly neutral shade to avoid any color conflict or interaction with simulated surface colors.

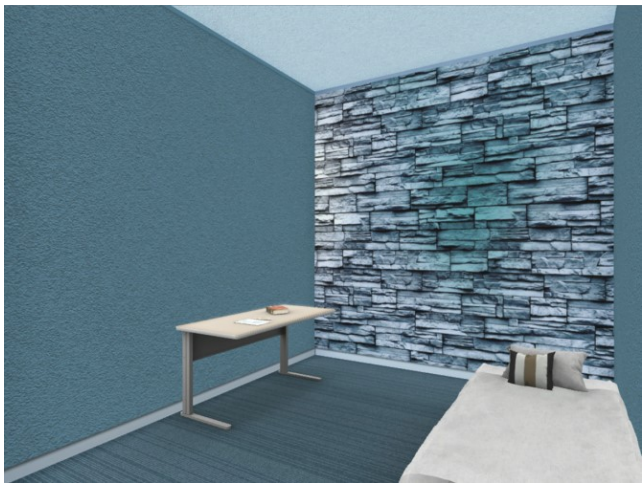

Figure 3: Texture mapping of simulated space—stone, drywall and fabric (carpet).

### 3.2.3 Simulating Design Parameters

The real-world color, light and texture (surface property of materials) of an interior space were represented by the color, brightness and texture gradient of texture maps of all six enclosing surfaces of the virtual living space:

a) Color—Orange (HSB 18,58,84) and blue (HSB 200, 48, 55); two complimentary hues were selected from the opposite spectrum of the color wheel.
b) Brightness—Two levels of brightness were created—bright and dark. Color tones of texture maps were used to represent real-world illumination. Increasing Saturation and lowering Brightness from the HSB (Hue, Saturation, Brightness) values of each color darkened the color tones. Light and dark color tones represented bright and dark brightness levels. The hue remained the same (figure 4).
c) Texture—Image-based texture maps of each material were modified to create two types of texture graininess for each material—rough and smooth (figure 5). Roughness in texture meant increased depth and sharpness of texture grains of a material. High resolution, photorealistic images of materials were manipulated in contrast, highlight and sharpness to create texture grains that closely represented rough and smooth characteristics of real surface properties of materials. Random variations for tiling, light and shadow were made for a more natural and realistic look.

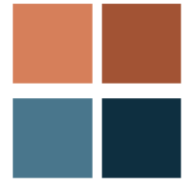

Figure 4: Color tones of orange and blue were darkened to represent two levels of brightness—bright (left) and dark (right).

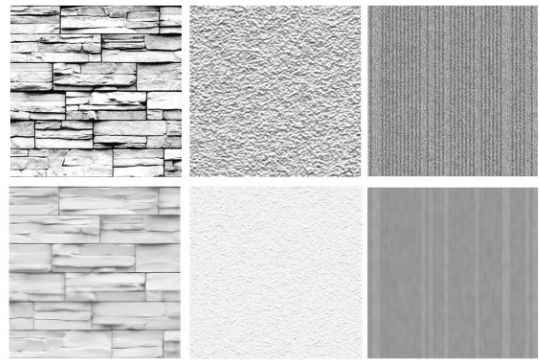

Figure 5: Image-based texture maps of stone, drywall and fabric (carpet) were modified to create two texture gradients—rough (top row) and smooth (bottom row).

### 3.2.4 Generating Virtual Test Spaces

Eight unique test spaces were created with two types of three design parameters, as shown in Table 2. Each space shared a unique combination of each design parameter (color, brightness and texture). The depiction of all test spaces is shown in figure 6.

Table 2: Each of the eight *test spaces* is composed of one characteristic of each design parameter.

| Design Parameters | | | Virtual Test Spaces |
|---|---|---|---|
| Color Type | Brightness Levels | Texture Gradients | |
| Orange | Dark | Rough | Orange – Dark – Rough |
| | | Smooth | Orange – Dark – Smooth |
| | Light | Rough | Orange – Light – Rough |
| | | Smooth | Orange – Light – Smooth |
| Blue | Dark | Rough | Blue – Dark – Rough |
| | | Smooth | Blue – Dark – Smooth |
| | Light | Rough | Blue – Light – Rough |
| | | Smooth | Blue – Light – Smooth |

The eight living space configurations were randomized and presented to users in the virtual environment for assessment. The order of display for the test spaces was randomized to minimize biases that could arise from tedium or familiarity. In each experimental group, each of the sixteen participants was allocated a unique randomized sequence. The randomization sequences were paired between groups.

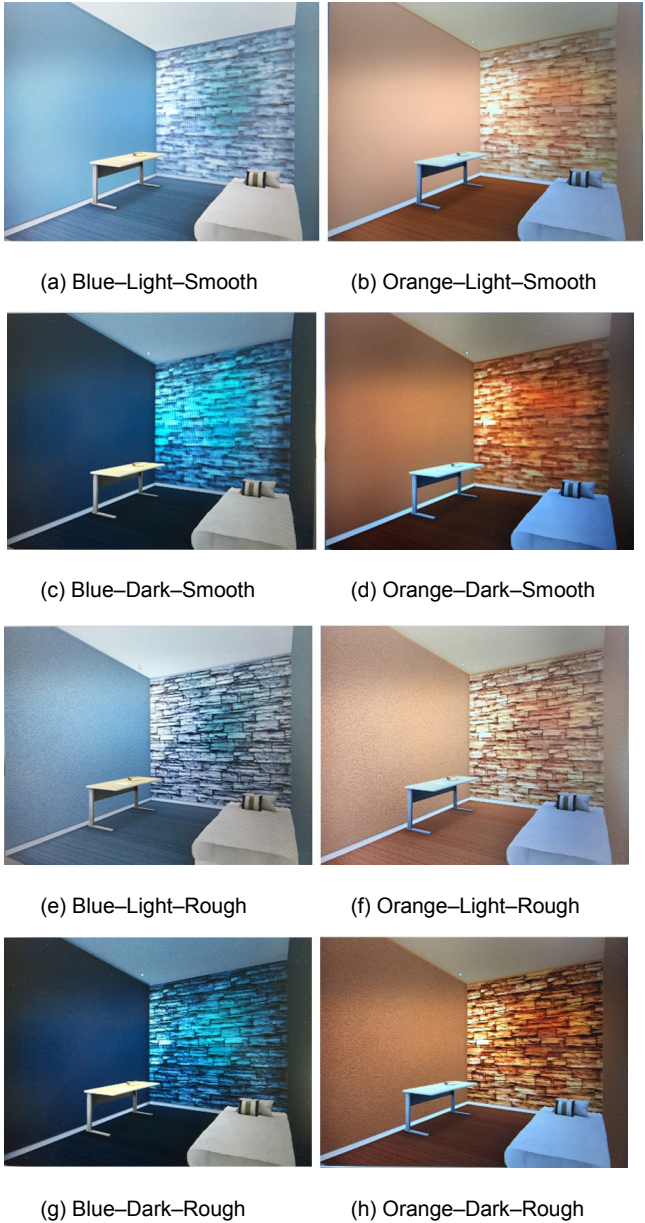

(a) Blue–Light–Smooth  (b) Orange–Light–Smooth
(c) Blue–Dark–Smooth  (d) Orange–Dark–Smooth
(e) Blue–Light–Rough  (f) Orange–Light–Rough
(g) Blue–Dark–Rough  (h) Orange–Dark–Rough

Figure 6: Screenshots of each of the eight *test spaces* composed of one characteristic of each design attribute.

### 3.2.5 Participants

A total of 32 participants were recruited for the study. Respondents were undergraduate, graduate and post-graduate students from various disciplines, as well as professionals (faculty and staff) at Duke University. Recruitment was done via email and poster announcements on the university campus. Participation was voluntary. Twenty-nine participants (91%) ranged in age from 18—40 (Mean =27, SD=4.8). Three participants (9%) were in the age range of 50—76. There were twenty-seven (84%) students and five (16%) professionals. Among the 27 students, five participants (16%) were in undergraduate level, thirteen (40%) were in master's level and fourteen (44%) were in doctoral or postdoctoral level.

### 3.2.6 Procedure

The whole experiment took approximately half an hour for each participant to complete. Participants were required to sign an informed consent form and fill out a demographic questionnaire before the test session commenced.

Participants were seated inside the virtual environment. Each of the eight test spaces were displayed in a pre-defined random order for each participant. Each space was displayed for 20 seconds before a set of questions was asked regarding that space. While inactive participants were allowed to stare at each space, active participants were asked to perform their task of folding clothes during those 20 seconds. They were asked to stop once the questions were asked. Their verbal responses were recorded.

### 3.2.7 Questionnaire

Participants were asked a set of questions verbally for each virtual test space displayed. The questions consisted of three categories of spatial dimensions used to measure the capacity of perceived spatial experience to influence moods or emotions, in terms of size, temperature and arousal level.

Participants were asked to rate each space in terms of specific psychophysiological aspects of space on a scale of 1 to 10. Here, 1 indicated the most negative or lowest rating and 10 indicating the most positive or highest rating. The aspects were: warm, cool, spacious, intimate, exciting, calm and comforting. They were also asked to rate each space for two activities: rest and work.

It was important to ensure that participants understood the semantics or contextual meaning of the spatial terms used in the questionnaire without being primed. Prior to the virtual session, the context of these psychophysiological aspects of spaces was explained to them in terms of pairs of oppositional adjectives: spacious and intimate, warm and cool, and exciting and calm. Participants were instructed to understand *exciting* spaces as the opposite of spaces that are *calming. Intimate* spaces were explained as the opposite of *spacious* spaces. *Cool* was described as a term opposite of *warm* to be understood in the context of eliciting a feeling of a warm or a cool temperature in the spaces.

## 4 RESULT

Data analysis was performed in IBM SPSS 24. A mixed-design factorial ANOVA with three repeated-measures factors and one between-subjects factor was performed. Data was visually inspected and verified to approach normality. Data is reported as statistically significant at p<.050. Statistically significant differences are denoted by an asterisk (*) in figures 7-12.

### 4.1 Main Effects

Figure 7 shows a summary of the main effects of color. Statistically significant main effects of color were found on perceptions of *warmth, coolness, excitement, calmness, intimacy, comfort,* and environments for *working* and *resting*. Orange was found significantly more *warm* ($F(1,30)=153.162$, $\eta^2=.836$), *exciting* ($F(1,30)=22.087$, $\eta^2=.424$) and *comfortable* ($F(1,30)=31.772$, $\eta^2=.514$) than blue at $p<.001$ for all measures. Orange was also found significantly more *intimate* than blue ($F(1,30)=4.960$, $p<.050$, $\eta^2=.142$), and significantly preferable for *resting* ($F(1,30)=10.732$, $\eta^2=.263$) and *working* ($F(1,30)=10.247$, $\eta^2=.255$) than blue both at $p<.005$. Blue was found to be significantly more *cool* ($F(1,30)=97.117$, $p<.001$, $\eta^2=.764$), and *calm* ($F(1,30)=4.918$, $p<.050$, $\eta^2=.141$) than orange.

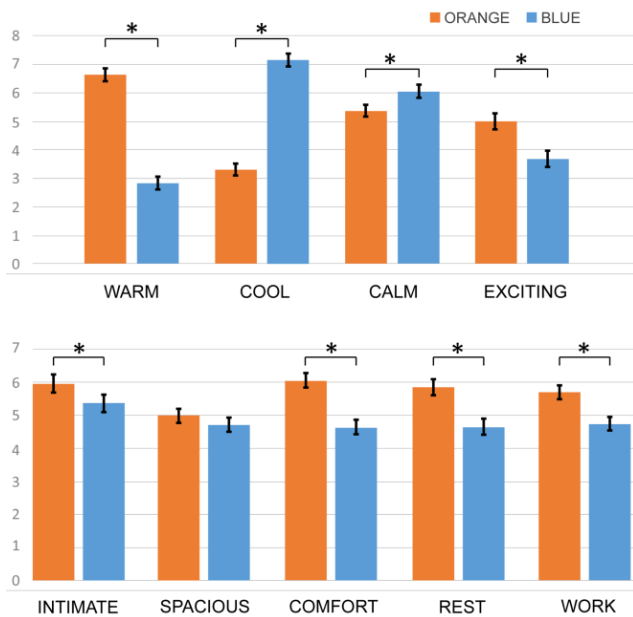

Figure 7: Main effects of color. Error bars represent standard errors.

A significant main effect of brightness was observed only on perception of *spaciousness*. As shown in Figure 8, bright space was found more *spacious* than dark space ($F(1,30)=5.615$, $p<.050$, $\eta^2=.158$). Also, a near-significant trend points in the direction that dark spaces were preferred over bright spaces for *resting* ($F(1,30)=4.169$, $p=.050$, $\eta^2=.263$).

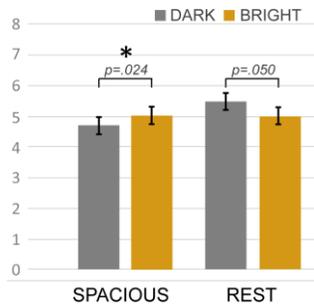

Figure 8: Brightness effects. Error bars represent standard errors.

No significant main effects of texture and activity were found on users' perceptions of emotional spaces. Although a strong trend towards significance ($p=.061$) was found that suggested that active participants perceived spaces more *intimate* than inactive participants did.

### 4.2 Interaction Effects

We found significant interaction effects between color and brightness on perceptions of *warmth* ($F(1,30)=11.262$, $p<.005$, $\eta^2=.273$); *coolness* ($F(1,30)=9.060$, $p<.050$, $\eta^2=.232$); *excitement* ($F(1,30)=7.922$, $p<.05$, $\eta^2=.209$), and *intimacy* ($F(1,30)=6.753$, $p<.050$, $\eta^2=.184$).

Pairwise comparisons show that for orange spaces, changing brightness made a significant difference to perception in specific situations. Orange felt significantly *warmer* and more *intimate* in dark environments, whereas orange felt significantly *cooler* in a bright environment (figure 9). On the contrary, blue felt significantly more *exciting* in bright environments, but significantly *calmer* in dark spaces. Orange felt significantly more *intimate* than blue only in dark spaces.

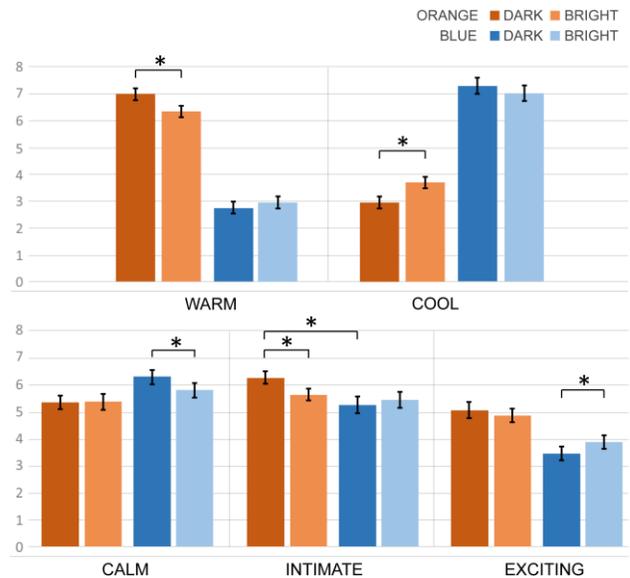

Figure 9: Interactions: Color and brightness. Error bars represent standard errors.

There was a significant interaction effect between brightness and texture on a participant's perception of environment preferable for *working* ($F(1,30)=4.815$, $p<.050$, $\eta^2=.138$). Pairwise comparisons show that in dark environments, smooth texture was found more preferable for *working* than rough texture, whereas no significant differences were found in bright spaces (figure 10).

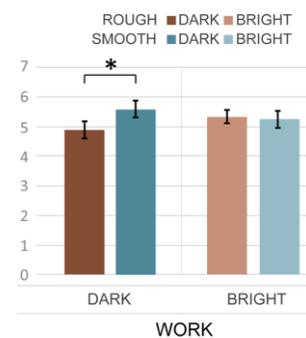

Figure 10: Interactions: Brightness and texture on environment for *working*. Error bars represent standard errors.

A significant interaction effect between color and activity was also found on the perception of *intimate* spaces ($F(1,30)=5.221$, $p<.050$, $\eta^2=.148$). For inactive participants, changes in color made a significant difference in the perception of *intimacy* of space, while no differences were observed for active participants. Inactive participants found orange significantly more *intimate* than blue, whereas no such significant difference was observed for active participants. Active participants found blue significantly more *intimate* than inactive participants did, while the same effect was not observed with orange spaces (figure 11).

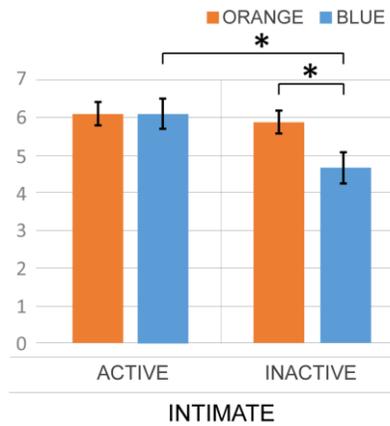

Figure 11: Interactions: Color and Activity on *intimacy* of space. Error bars represent standard errors.

Significant interaction effect of brightness and activity was found only on the perception of *warmth* ($F(1,30)=4.395$, $p<.050$, $\eta^2=.128$). For active participants, changes in level of brightness made significant difference in the perception of *warmth*. Active participants rated dark spaces more *warm* than bright spaces. No such significant difference was observed for inactive participants (figure 12).

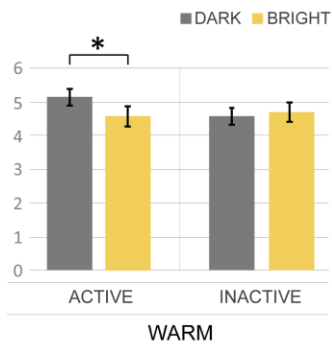

Figure 12: Interactions: Brightness and activity on *warmth* of space. Error bars represent standard errors.

Finally we found a significant four-way interaction effect among color, texture, brightness and activity on perception of *intimacy* of space ($F(1,30)= 4.651$, $p<.050$, $\eta^2=.134$).

## 5 Discussion

Comparisons were drawn between pairs of oppositional adjectives—warm and cool, exciting and calm, spacious and intimate—to find out if perceptions of spatial aspects represented by oppositional adjectives elicit contrasting emotional responses. Data revealed that orange was found significantly more *warm* and *exciting* than blue, whereas blue was found significantly more *cool* and *calming* than orange, regardless of level of brightness. It can be deduced from this data that spatial qualities represented by oppositional adjectives evoked contrasting emotional responses. This data provides evidence that the VE can generate spatial aspects of *warmth, coolness,* and *excitement* and *calmness* that were perceivable to participants. Thus, the hypothesis posed at the beginning of this paper as a basis for this study can be confirmed.

However, comparisons of data between oppositional adjectives *spacious* and *intimate* reveal a different result. While color had a significant impact on the perception of *intimate* spaces, it did not have any significant impact on perception of *spaciousness*. In fact, brightness was found to have a significant effect on perceived spaciousness of *space*. From this data, it can be concluded that the participants chose to consider a different or broader semantic for the word *intimate* in their understanding of space. It is possible that they did not perceive the term *intimate* as a directly oppositional adjective for the term *spaciousness* of spatial experience.

The study was able to replicate certain real-world perceptions related to color and brightness in the virtual environment. In real-world color perception, warm colors are generally found more exciting than cool colors, whereas cool colors are found more calming than warm colors. Also, in real-world perception of space, warm colors appear to be closer in distance than cool colors. The data gathered in this virtual test revealed that orange was perceived *warmer,* more *exciting* and *intimate* than blue. On the other hand, blue was perceived more *calm* and *cool* than orange. This correlated with real-world color experience. Additionally, the user study could also replicate the real-world phenomenon of brightness having a significant impact on perception of *spaciousness*.

## 6 Formulating Design Principles

The role of VR in the aesthetic practice of space design is the use of "sensory substitution methods" to replicate the multi-sensorial real world by understanding how to override real-world sensory perceptions and in that process, provide affordance for a new aesthetics that conforms to the medium [18]. In this study, the real-world spatial experiences that are essentially multi-sensory and three-dimensional are simulated within the technological constraints of VR as visually perceptible experiences with equivalent graphical methods. The study quantifies the aesthetic parameters of architectural space-making that are pertinent to design. The underlying aesthetics of a medium influences its capacity and efficiency [18]. This fundamental aesthetics affects VR's effectiveness in allowing compelling space creation that triggers human emotional response and potential activeness through perception of sensorial spatial elements.

A set of design guidelines has been developed as the underlying aesthetics for space design in which one or more parameters can be modified to create desired spatial qualities with affective dimensions. These principles can be used by architects and interior designers as a foundation for living space design for *neo-nomadic* minimalist lifestyles.

For design principles, the user study findings of established correlations between design parameters and perceptions of psychophysiological spatial aspects are used as a basis. Three categories of psychophysiological spatial aspects have been selected, each with a pair of oppositional affective dimensions: temperature (warm and cool), size (spacious and intimate) and level of arousal (excitement and calmness). As it was found in the study that spatial qualities represented by oppositional adjectives evoked contrasting emotional responses, inferences could be drawn to compare each pair to establish the design principles. For example, data for *cool* and *calm* spaces could be inferred from relevant data found for *warm* and *exciting* spaces respectively. Relevant data for *intimate* spaces can also be inferred from data on *spaciousness*. Here, the term *intimate* is considered as an exact opposite adjective of the term *spaciousness*.

For surface color modifications of space, a color spectrum or a traditional color wheel is used to identify ranges of warmer or cooler shades of any given color [1]. Here, pure red is considered the warmest color and pure blue is considered the coolest color. Although the study presented only two colors—orange and blue— the design guidelines are formed with the assumption that orange represents warm colors within the range of red, orange and yellow on the warmer side of the color spectrum, and blue represents cool colors within the range of green, blue and purple on the cooler side of the color spectrum.

Table 3 shows the design principles formulated for an occupant's desired affective spatial qualities. Statistically significant correlations formed guidelines mainly for two design parameters—color and brightness. Rules are also formulated for active occupants. Color is the primary design parameter that can be modified to create feelings of *warmth, coolness, excitement* and *calmness*. For *warmer* or more *exciting* spaces, a warmer shade of warm colors can be used. For any given color, its warmer shades are its adjacent colors that are located closer to red in the color wheel. For example, yellowish-orange is a warmer color than yellow. On the other hand, for *cooler* and *calmer* spaces, cooler shades of cool colors can be used. For example, greenish-blue is a cooler color than green.

In addition to color, brightness levels can also be modified as the secondary design parameter to further manipulate these feelings. For a space that has warm colors, decreasing brightness level can make it feel *warmer*, and increasing it can make it feel *cooler*. On the other hand, for a space that has cool colors, brightness level can be increased to make it feel more *exciting*, and decreased to make it feel *calmer*.

As a primary design parameter, brightness levels can be modified to increase or decrease feeling of *spaciousness*, and by inference, the feeling of *intimacy*.

Table 3: Correlations between user's desired spatial qualities and corresponding design principles

| Desired Affective Aspects | Design Parameters to modify | Design Principles |
|---|---|---|
| Warm | Color Brightness | • Use warm colors—variations of red, orange or yellow<br>• Use warmer shades of warm colors<br>• If existing color of space is warm, decrease brightness to make it warmer<br>• For active occupants, decrease brightness |
| Cool | Color Brightness | • Use cool colors—variations of blue, green or purple<br>• Use cooler shades of cool colors<br>• If occupant prefers warm colors, increase brightness to make it cooler<br>• For active occupants, increase brightness |
| Spacious | Brightness | • Increase brightness |
| Intimate | Brightness | • Decrease brightness |
| Exciting | Color Brightness | • Use warm colors—variations of red, orange or yellow<br>• Use warmer shades of warm colors<br>• If occupant prefers cool colors, increase brightness |
| Calm | Color Brightness | • Use cool colors—variations of blue, green or purple<br>• Use cooler shades of cool colors<br>• If room already has cool colors, decrease brightness |

## 7 Conclusion & future work

The outcome of the study assisted in formulating aesthetic guidelines for affective living spaces pertaining to sensory engagement. One major focus of future work is to explore architectural design elements that can potentially evoke feelings of spaciousness in a tiny, shared living space intended for young professionals. It includes virtually simulated sunlight and openings (windows) as design elements to examine their affective dimensions. Perception of interior surfaces can extend beyond the traditional wall and reflect abstract representations of external views, weather conditions, and real-time diurnal and seasonal changes with the use of simulated color, light and texture.

This research can be taken further into the domain of sensor-driven interactive living spaces. Interactive architecture is defined as the ability to provide variable characteristics of living, working and leisure spaces. While VR can serve as an architectural evaluation tool to test, replicate and evaluate real-world spatial situations, its interactive capacity can be utilized to explore real-time space-making as an aesthetic medium. The use of VEs can provide a highly effective aesthetic tool that allows changing of spaces to trigger human emotional responses in order to support human activities through interaction. The immersive human-scale CAVE-type display can work as an architectural design tool to create a variety of spaces for testing and applying in real-world interactive living spaces.

## Acknowledgments

Experiments at the Duke University DiVE were made possible by a Dissertation Research Award from the University of Texas at Dallas and by Duke University. Dr. Timothy J. Senior (Neuroscience, Oxford University) suggested DiVE as an immersive environment for experimental evidence of design choices. Hypotheses informing the experiments were presented at the international conference "Anticipation Across Disciplines" (September 2014). The presentation was made possible through funding from the Hanse Institute for Advanced Study (Hanse Wissenschaftskolleg, Germany) and the antÉ – Institute for Research in Anticipatory Systems. The inter-institutional IRB is the result of cooperation between the University of Texas and Duke University.